\documentclass[%
 reprint,
showpacs,
 amsmath,amssymb,
 aps,
prc
]{revtex4-1}

\usepackage{graphicx}
\usepackage{dcolumn}
\usepackage{bm}

\begin{document}


\title{Phenomenology of \(f_2(1270)\) photoproduction at energies measured with the CLAS facility
}

\author{K. E. S. Mendes}
\affiliation{Instituto de F\'{\i}sica e Matem\'atica, Universidade Federal de Pelotas, \\ Caixa Postal 354, CEP 96010-090, Pelotas, RS, Brazil.}

\author{D. T da Silva}
\affiliation{Instituto de F\'{\i}sica e Matem\'atica, Universidade Federal de Pelotas, \\ Caixa Postal 354, CEP 96010-090, Pelotas, RS, Brazil.}

\author{M. L. L. da Silva}
\email{mllsilva@gmail.com}
\affiliation{Instituto de F\'{\i}sica e Matem\'atica, Universidade Federal de Pelotas, \\ Caixa Postal 354, CEP 96010-090, Pelotas, RS, Brazil.}%


\date{\today}

\begin{abstract}
We investigate the photoproduction of the tensor meson $f_2(1270)$ on the proton within a
Regge-based framework, focusing on the reaction $\gamma p \to p f_2(1270)$ in the few-GeV
energy region. The production mechanism is modeled through the exchange of vector-meson
Regge trajectories in the $t$-channel, including both $\rho$ and $\omega$ exchanges with
phenomenologically motivated couplings. The scattering amplitudes are derived using Reggeized
propagators and effective hadronic vertices, allowing for the calculation of differential
cross sections in the narrow-width approximation. We further extend the analysis to the
$\pi^+\pi^-$ invariant mass distribution by incorporating a relativistic Breit--Wigner
description of the $f_2(1270)$ resonance.
\end{abstract}

\maketitle


\section{Introduction}

The photoproduction of mesons off the nucleon provides a powerful tool for investigating the mechanisms
governing strong interactions in the non-perturbative regime of QCD. In particular, exclusive reactions
induced by real photons allow one to probe the interplay between hadronic degrees of freedom, exchange 
mechanisms, and resonance dynamics over a wide range of energies \cite{Laget:2019tou}. In the few-GeV
energy domain, where perturbative methods are not applicable, phenomenological approaches
based on effective hadronic models and Regge theory have proven especially successful
\cite{Battaglieri:2014gca, Holinde, Donnachie2002}. Within this context, the study of tensor-meson
photoproduction offers a valuable opportunity to test our understanding of meson–baryon interactions
beyond the well-explored pseudoscalar and vector meson sectors.

The $f_2(1270)$ meson is the lightest isoscalar tensor meson, with quantum numbers $J^{PC}=2^{++}$, mass
$M = 1275.4\pm0.8 \,\text{MeV}$, width $\Gamma_{\rm Tot} = 185.8_{-2.1}^{+2.8}\, \text{MeV}$ and is
commonly interpreted as a predominantly non-strange $q\bar{q}$ state \cite{ParticleDataGroup:2024cfk,
CDK, Giacosa:2009qh}. Owing to its strong coupling to the $\pi\pi$ channel, the $f_2(1270)$ plays a
central role in two-pion dynamics and has been extensively studied in hadronic reactions and decays.
Despite this, its production mechanisms in electromagnetic processes are comparatively less constrained,
particularly in exclusive photoproduction reactions. Understanding the dynamics underlying the reaction
$\gamma p \to p\,f_2(1270)$ therefore provides important insight into the coupling of tensor mesons to
photons and vector mesons, as well as into the role of spin and angular momentum in hadronic production
processes.

Photoproduction of mesons at photon energies of a few GeV and above is dominated by $t$-channel
exchange mechanisms, leading to characteristic forward-peaked angular distributions. In this regime,
reactions such as $\gamma p \to p\,\pi\pi$ can be described in terms of meson exchanges that
subsequently materialize into resonant states, including scalar, vector, and tensor mesons
\cite{CLAS:2009ngd,CLAS:2020ngl, daSilva:2011hy,daSilva:2013yka, daSilva:2014xya}. The extraction
of resonance contributions from multipion final states has become increasingly precise with the
advent of high-statistics experiments, enabling detailed studies of differential cross sections and
partial-wave content \cite{Battaglieri:2014gca}. These developments motivate dedicated theoretical
investigations of specific resonant channels, such as $f_2(1270)$ photoproduction, within a
consistent dynamical framework \cite{Mathieu:2020zpm, Xie:2014twa}.

Regge theory provides a natural and well-established description of high-energy hadronic reactions
dominated by $t$-channel exchanges. By replacing individual Feynman propagators with Reggeized
trajectories, this approach effectively accounts for the exchange of entire families of particles
sharing the same quantum numbers \cite{Donnachie2002}. For photoproduction processes, Regge models
based on vector-meson exchange have successfully described a wide variety of reactions, reproducing
both the energy dependence and the forward-angle behavior of differential cross sections. In the
case of tensor-meson photoproduction, the exchange of $\rho$ and $\omega$ trajectories is expected
to play a dominant role, making Regge theory a particularly suitable framework for modeling the
reaction $\gamma p \to p\,f_2(1270)$.

On the experimental side, recent measurements by the CLAS Collaboration at Jefferson Lab have
provided the first dedicated data on $f_2(1270)$ photoproduction, extracted from the reaction
$\gamma p \to p\,\pi\pi$ through partial-wave analysis \cite{CLAS:2009ngd, CLAS:2020ngl}. These
results cover photon energies in the few-GeV range and exhibit a pronounced forward peaking
characteristic of $t$-channel exchange dynamics. In parallel, the GlueX experiment is producing
high-precision photoproduction data with polarized photon beams, offering access to polarization
observables and interference effects that are particularly sensitive to the underlying production
mechanisms \cite{Accardi:2023chb}. Together, these experimental advances call for theoretical
models capable of providing a unified description of existing data and reliable predictions for
future measurements \cite{Battaglieri:2014gca}.

In this work, we investigate the photoproduction of the $f_2(1270)$ meson on the proton within a
Regge-based framework, focusing on vector-meson exchange in the $t$-channel. We derive the
corresponding scattering amplitudes and differential cross sections, analyze their energy and
momentum-transfer dependence, and compare our results with available CLAS data. Our approach aims
to provide a transparent and phenomenologically grounded description of tensor-meson
photoproduction and to establish a baseline for future experimental and theoretical studies in this
sector.

\section{Model and Cross Section Calculation}

The Regge-based description adopted here is expected to be valid in the high-energy, forward-angle
regime, where the Mandelstam variable (s) is large compared to the hadronic mass scales and the
momentum transfer (t) remains small and negative. In this kinematic region, the reaction dynamics
is dominated by (t)-channel exchanges, and the use of Reggeized vector-meson propagators
effectively resums the contributions from entire families of exchanged states lying on the
corresponding $\rho$ and $\omega$ trajectories.

The reaction under consideration is
$
\gamma p \rightarrow p f_2(1270).
$
Within the framework of Regge theory, the differential cross section in the narrow-width limit
for a tensor meson of mass $m_T$ is given by
\begin{eqnarray}
\frac{d\hat{\sigma}}{dt}(\gamma p \rightarrow p f_2(1270)) =
\frac{|{\cal M}(s,t)|^2}{64\pi (s-m_p^2)^2},
\label{dsigma}
\end{eqnarray}
where ${\cal M}$ denotes the scattering amplitude, $s$ and $t$ are the usual Mandelstam variables,
and $m_p$ is the proton mass.

For the exchange of a single vector meson ($\rho$ or $\omega$), the squared invariant amplitude
can be written as
\begin{eqnarray}
|{\cal M}(s,t)|^2 &=&
-\frac{1}{2}{\cal A}^2(s,t)\Big[
s(t-t_1)(t-t_2) \nonumber \\
&& + \frac{1}{2}t\big(t^2 - 2(m_T^2+s)t + m_T^4\big)
  \Big] \nonumber\\
  &-& {\cal A}(s,t) {\cal B}(s,t) m_p s(t-t_1)(t-t_2) \nonumber\\
  &-& \frac{1}{8} {\cal B}^2(s,t) s(4m_p^2-t)(t-t_1)(t-t_2),
  \label{msquare}
\end{eqnarray}
where $t_1$ and $t_2$ are the kinematical limits of the momentum transfer,
\begin{eqnarray}
  t_{1,2} &=&
  \frac{1}{2s}\Big[
  -(m_p^2-s)^2 + m_T^2(m_p^2+s)
  \nonumber\\
  &\pm&
  (m_p^2-s)\sqrt{(m_p^2-s)^2
  -2m_T^2(m_p^2+s)+m_T^4}
  \Big].
  \label{t12}
\end{eqnarray}

The structure of the squared amplitude in Eq.~(\ref{msquare}) reflects the interplay between
the helicity-conserving and helicity-flip components of the $VNN$ vertex, encoded in the
functions ${\cal A}(s,t)$ and ${\cal B}(s,t)$, respectively. The factors $(t-t_1)(t-t_2)$
enforce the correct kinematical constraints and ensure that the amplitude vanishes at the
physical boundaries of the momentum transfer. In the forward limit, where $|t|$ is small, the
contribution proportional to ${\cal A}^2(s,t)$ is expected to dominate, while terms involving
${\cal B}(s,t)$ are suppressed by powers of $t$ and the proton mass.

The Reggeization of the Feynman propagators is performed assuming linear Regge trajectories,
$\alpha_V(t) = \alpha_{V}(0) + \alpha'_V t$, which leads to the following expressions for the
Regge amplitudes ${\cal A}(s,t)$ and ${\cal B}(s,t)$:
\begin{eqnarray}
{\cal A}_V (s,t) &=&
g_A\left(\frac{s}{s_0}\right)^{\alpha_V(t)-1}\!\!\!\!
\frac{\pi\alpha'_V}{\sin(\pi\alpha_V(t))\Gamma(\alpha_V(t))}
D_V(t), \nonumber\\
{\cal B}_V (s,t) &=&
-\frac{g_B}{g_A}{\cal A}_\rho (s,t),
\label{abdefrho}
\end{eqnarray}
where $D_V(t)$ is the signature factor. Degenerate $\rho$ trajectories are assumed
\begin{eqnarray}
 D_V(t) = exp(-i \pi \alpha_\rho(t)).
\end{eqnarray}
This approximation is typically valid in the high-energy regime. In contrast, the $\omega$
trajectory is treated as non-degenerate
\begin{eqnarray}
 D_V(t) = \frac{-1 + exp(-i \pi \alpha_\omega(t))}{2}.
\end{eqnarray}
This choice is phenomenologically motivated, as it allows interference effects between
$\rho$ and $\omega$ exchanges to contribute to the shape and normalization of the
differential cross section \cite{Laget:2019tou,Kochelev:2009xz}. The $\rho$ and $\omega$
trajectories with intercepts and slopes are given by
$
\alpha_V(0)=0.55 (0.44)$ and $\alpha'_V = 0.8 (0.9)\text{GeV}^{-2},
$
for $\rho$ ($\omega$) exchange, respectively.

The relative importance of $\rho$ and $\omega$ exchanges is governed both by their Regge
intercepts and by the corresponding coupling constants. Owing to its larger intercept,
the $\rho$ trajectory provides the leading contribution at high energies, whereas
the $\omega$ exchange, although subleading, can still play a non-negligible role through
interference effects, particularly in polarization observables and in the intermediate
energy region. The different signature factors in Eq.~(\ref{abdefrho}) encode the
naturality properties of the exchanged trajectories and lead to characteristic phase
differences between the $\rho$ and $\omega$ amplitudes.

The couplings appearing in Eq.~(\ref{abdefrho}) are defined as
$
g_A = g_{\gamma VT}(g_V + 2m_p g_T)$ and $g_B = 2 g_{\gamma VT} g_T,
$
where $g_V$ and $g_T$ denote the vector and tensor couplings for the $VNN$ vertex,
respectively, and $g_{\gamma VT}$ is the coupling for the $\gamma VT$ vertex.
The $\omega NN$ couplings are relatively well determined~\cite{Holinde}, and we adopt
$g_V^\omega = 15$ and $g_T^\omega = 0$, following Ref.~\cite{DK}. For the $\rho NN$
couplings, which are less constrained, we use $g_V^\rho = 3.4$ and
$g_T^\rho = 11~\text{GeV}^{-1}$ \cite{DK}.
The $\gamma VT$ coupling $g_{\gamma VT}$ is treated as a free parameter. However, it is
constrained by the relation
$
g_{\gamma \rho f_2} = 3 g_{\gamma \omega f_2},
$
as motivated by Ref.~\cite{Mathieu:2020zpm}. It reflects the underlying flavor structure
of the $f_2(1270)$ tensor meson and provides a phenomenological constraint that reduces
the number of free parameters in the model. Within this framework, the present approach
allows for a consistent description of the energy and momentum-transfer dependence of
the $\gamma p \to p f_2(1270)$ cross section and provides a natural basis for comparison
with existing and future photoproduction data \cite{CLAS:2020ngl}.

The $D$-wave contribution to the nondiffractive photoproduction of the $f_2(1270)$ meson
in the $\pi^+\pi^-$ final state is considered. According to the Regge phenomenology, one
expects that only the $t$-channel meson exchanges are important in such a case. The $\rho$
and $\omega$ reggeized exchanges are to be considered in the present analysis. To obtain
mass distribution for the tensor $f_2(1270)$ meson, one represents it as relativistic
Breit--Wigner resonance with energy-dependent partial width. The differential cross section
for the production of a tensor with invariant mass $M$, and its decay to two pseudoscalars
can be written as,
\begin{eqnarray}
\frac{d\sigma}{dt~dM}=\frac{d\hat{\sigma}(t,m_T)}{dt}\frac{2m_T^2}{\pi}
\frac{\Gamma_i(M)}{(m_T^2-M^2)^2+ (M\Gamma_{\rm Tot})^2} ,
\label{signaldcs}
\end{eqnarray}
where $d\hat{\sigma}/dt$ is the narrow-width differential cross section in Eq.~(\ref{dsigma})
at a tensor mass $M=m_T$ and where $\Gamma_i(M)$ denotes the partial width for the
pseudoscalar--pseudoscalar final state. The $\Gamma_{\pi^+\pi^-}(M)$ is extracted from the
experimental value given by
\begin{eqnarray}
 \frac{\Gamma_{\pi^+\pi^-}}{\Gamma_{\rm Tot}} = 0.562^{+0.019}_{-0.006}
\end{eqnarray}
and the total width of $f_2(1270)$ \cite{ParticleDataGroup:2024cfk}. The present model is
expected to be reliable in the high-energy, forward-angle regime, where Regge-pole dominance
is justified.

\section{Results}

In this section, we present the numerical results for the differential cross section of the reaction
$\gamma p \to p\,f_2(1270)$, obtained within the Regge-based framework described in the previous section.
Our calculations focus on photon energies corresponding to the kinematic range covered by the CLAS
experiment, allowing for a direct comparison with available experimental data.

\begin{figure}[ht]
\begin{center}
\includegraphics[scale=0.34]{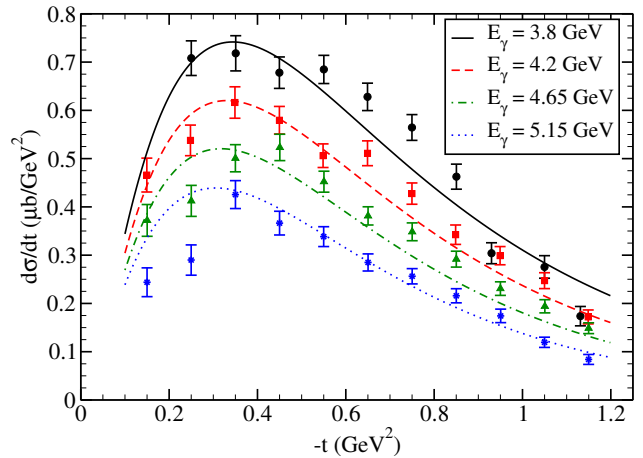}
\caption{
Differential cross section for the reaction $\gamma p \to p f_2(1270)$ as a function of $-t$ for
distinct photon energies. The curves represent the results of the Regge model including $\rho$
and $\omega$ exchanges. The data points correspond to CLAS measurements \cite{CLAS:2020ngl}.
}
\label{diff_cross}
\end{center}
\end{figure}

Fig.~\ref{diff_cross} shows the differential cross section $d\hat{\sigma}/dt$ as a function of the
momentum transfer $-t$ for four representative photon energies, $E_\gamma = 3.8$ , $4.2$, $4.65$, and
$5.15$~GeV and the free parameter $g_{\gamma\rho f_2} = 0.37\,\, {\rm GeV^{-1}}$. The results exhibit
a pronounced forward peaking, characteristic of reactions dominated by
$t$-channel exchange mechanisms. This behavior is a natural consequence of the Reggeized vector-meson
exchange, where the energy dependence is governed by the factor $(s/s_0)^{\alpha_V(t)-1}$ and the
shape of the $t$-distribution is controlled by the trajectory slope $\alpha'_V$.

As shown in the Fig.~\ref{diff_cross}, the model reproduces well the overall magnitude and the slope
of the experimental data in the forward region. The exponential fall-off of the cross section with
increasing $-t$ is consistent with the exchange of $\rho$ and $\omega$ Regge trajectories and reflects
the dominance of helicity-conserving amplitudes at small momentum transfer. In this region, the
contribution proportional to ${\cal A}^2(s,t)$ provides the leading behavior, while terms involving
${\cal B}(s,t)$ give subleading corrections.

The energy dependence of the cross section is also well described by the model. As the photon energy
increases, the forward cross section shows a moderate rise, in agreement with the intercepts of the
vector-meson trajectories employed in the calculation. The dominance of the $\rho$-trajectory, due to
its larger intercept, is evident at higher energies, while the $\omega$-exchange contributes mainly
through interference effects and affects the overall normalization.

The normalization of the calculated cross sections is fixed by the $\gamma VT$ coupling constants.
The phenomenological constraint $g_{\gamma\rho f_2} = 3\,g_{\gamma\omega f_2}$, motivated by
flavor symmetry considerations, plays an essential role in reducing the number of free parameters
and leads to a consistent description of the data across the entire energy range considered.
Within the present uncertainties, no additional mechanisms beyond vector-meson exchange are
required to account for the CLAS measurements in the forward region.

Deviations between the model and the data at larger values of $-t$, where present, may signal the
onset of additional contributions not included in the present approach, such as Regge cuts,
rescattering effects, or the breakdown of the simple Regge-pole dominance.
Nevertheless, in the kinematic region where Regge theory is expected to be applicable, the agreement
with the data supports the validity of the present description.

Overall, the results demonstrate that tensor-meson photoproduction in the few-GeV energy region can
be consistently described within a Regge framework based on vector-meson exchange.
The present calculation provides a reliable baseline for future studies, including extensions to
polarization observables and comparisons with forthcoming high-precision data from the GlueX
experiment.

\begin{figure}[ht]
\begin{center}
\includegraphics[scale=0.34]{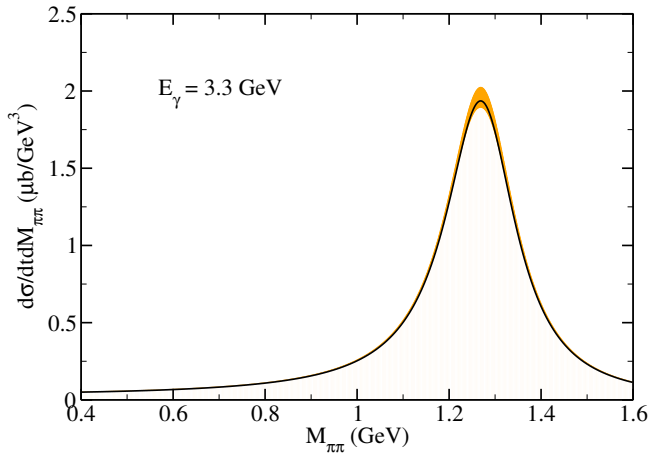}
\caption{
Invariant mass distribution for the reaction $\gamma p \to p \pi^+ \pi^-$ as a function of the
invariant mass $M$ of the $\pi^+\pi^-$ system at $E_\gamma = 3.3$ GeV and fixed momentum
transfer $-t = 0.55$ GeV$^2$.
}
\label{mass_dist}
\end{center}
\end{figure}

The invariant mass distribution shown in Fig.~\ref{mass_dist} provides further insight into the
production and decay dynamics of the $f_2(1270)$ meson. The characteristic resonant peak around
$M \approx 1.27$ GeV reflects the dominance of the tensor meson contribution in the $\pi^+\pi^-$
channel and the width of the distribution is compatible with the experimental value within the
expected resolution of the model~\cite{CLAS:2009ngd}. The shape of the distribution is well
described by a relativistic Breit--Wigner form, modulated by the energy-dependent partial width
associated with the $D$-wave decay of the tensor meson.

The overall normalization of the distribution is directly connected to the underlying
photoproduction mechanism encoded in the Regge amplitude. In particular, the $t$-channel exchange
of vector mesons determines the production strength, while the decay dynamics controls the shape
in $M$. The factorization observed in Eq.~(\ref{signaldcs}) allows for a transparent separation
between production and decay processes.

Deviations from the pure Breit--Wigner shape, which are not included in the present model, may
arise from non-resonant $\pi\pi$ backgrounds, interference with other partial waves, or final-state
interactions. Such effects could become relevant in a more refined analysis and in comparisons with
high-precision experimental data.

\section{Conclusions}

In this work, we have investigated the photoproduction of the tensor meson $f_2(1270)$ on the
proton within a Regge-theory framework, focusing on the reaction $\gamma p \to p f_2(1270)$ in
the few-GeV energy region. The production mechanism was modeled through the exchange of
vector-meson Regge trajectories in the $t$-channel, incorporating both $\rho$ and $\omega$
exchanges with phenomenologically motivated couplings.

Using Reggeized propagators and effective hadronic vertices, we derived the corresponding
scattering amplitudes and evaluated the differential cross section in the narrow-width
approximation. The resulting $t$- and energy-dependence of the cross section exhibits the
characteristic forward peaking expected from $t$-channel dominated dynamics. Our calculations
provide a good description of the available CLAS data for several photon energies, reproducing
both the slope and the overall normalization of the measured differential cross sections in the
kinematic region where Regge theory is expected to be applicable.

The results indicate that the dominant contribution to $f_2(1270)$ photoproduction arises from
$\rho$-trajectory exchange, driven by its larger Regge intercept, while the $\omega$ exchange
plays a secondary but non-negligible role, mainly through interference effects. The
phenomenological constraint on the $\gamma V T$ couplings,
$g_{\gamma \rho f_2} = 3 g_{\gamma \omega f_2}$, significantly reduces the number of free
parameters and leads to a consistent description of the data across the energy range considered.

In addition to the differential cross section, we have analyzed the $\pi^+\pi^-$ invariant mass
distribution by incorporating a relativistic Breit--Wigner description of the $f_2(1270)$
resonance with an energy-dependent width. The resulting mass spectrum exhibits a clear resonant
peak around $M \approx 1.27$ GeV, consistent with the dominance of the tensor meson contribution
in the $\pi\pi$ channel. The shape of the distribution reflects the interplay between the
production mechanism, governed by Regge dynamics, and the decay process, controlled by the
$D$-wave nature of the resonance. A quantitative comparison with experimental invariant mass
spectra would require the inclusion of non-resonant backgrounds and detector acceptance effects.

We note that the invariant mass distribution obtained in this work is in good qualitative
agreement with the experimental results reported by the CLAS Collaboration, in particular with
Fig.~14 of Ref.~\cite{CLAS:2009ngd}. The position and width of the $f_2(1270)$ peak, as well
as the overall shape of the $\pi^+\pi^-$ spectrum, are consistently reproduced within the
present framework. This agreement indicates that the model captures the essential features of
both the production dynamics and the subsequent decay into the $\pi\pi$ channel.

The factorized form of the double differential cross section allows for a transparent separation
between production and decay dynamics. While the $t$-channel exchange determines the overall
normalization and energy dependence, the invariant mass distribution is primarily sensitive to
the resonance properties and phase-space effects. Deviations from the pure Breit--Wigner
behavior may arise from non-resonant backgrounds, interference with other partial waves, or
final-state interactions, which are not included in the present approach.

Deviations observed at larger values of the momentum transfer may point to the presence of
additional mechanisms not included in the present model, such as Regge cuts, rescattering
effects, or limitations of simple Regge-pole dominance. Nevertheless, within the forward-angle
regime, the agreement between the model and the experimental data supports the validity of a
Regge-based description for tensor-meson photoproduction.

The present study provides a consistent and phenomenologically grounded description of both
the production and decay aspects of $f_2(1270)$ photoproduction. It establishes a solid
baseline for future investigations, including extensions to polarization observables, a more
detailed treatment of non-resonant contributions, and comparisons with forthcoming
high-precision data from the GlueX experiment.

\begin{acknowledgments}
The authors thank Magno V. T. Machado, Rafael Cavagnoli, and Wagner Tenfen for valuable
comments and suggestions. This study was financed in part by the Coordena\c c\~ao de
Aperfei\c coamento de Pessoal de N\'{\i}vel Superior - Brasil (CAPES) - Finance Code 001.
\end{acknowledgments}




\end{document}